\documentclass[twoside,3p,times,twocolumn]{elsarticle}
\usepackage{epsfig}
\usepackage{amssymb}

\begin{document}

\begin{frontmatter}

\title{Measurement of the longitudinal spin structure of the proton by COMPASS\vspace*{-0.3cm}}

\vspace{-1cm}

\author{A. Korzenev\corref{cor1}}
\ead{korzenev@mail.cern.ch}
\author{for the COMPASS collaboration}
\cortext[cor1]{on leave of absence from JINR, 141980 Dubna, Russia}
\address{CEA Saclay, IRFU/SPhN, 91191 Gif-sur-Yvette, France\vspace*{0.2cm}
  \\
{\normalsize {\sf Talk was given at the 3rd joint International HADRON STRUCTURE '09 Conference,\\
 August 30 - September 3, 2009, Tatransk\'{a} \v{S}trba (Slovak Republic)}}\vspace*{-1cm}}

\begin{abstract}
The inclusive $A_{1,p}$ and hadron double-spin asymmetries 
$A^{\pi+}_p$, $A^{\pi-}_p$, $A^{K+}_p$, $A^{K-}_p$ measured         
at COMPASS (CERN SPS) in deep-inelastic scattering of a polarized
muon beam off a polarized NH$_3$ solid target are presented.
The results have been obtained with the full statistics collected
in 2007 for the longitudinal target polarization.
Proton asymmetries have been combined with the published deuteron ones.
An evaluation of the non-singlet spin-dependent structure function
$g_1^{NS}(x,Q^2)$ and its first moment,
which confirms the validity of the Bjorken sum-rule, is presented. 
A LO evaluation of polarized quark densities is also presented. 
The use of the proton data allows to perform a full flavor separation 
and to extract individual helicity densities of $u$, $d$, $\bar{u}$, 
$\bar{d}$ and $s$ quarks. All sea quark densities are found to be 
compatible with zero in the full range of the measurements.
\end{abstract}

\end{frontmatter}


\section{Introduction}

New results of the COMPASS experiment at CERN on the proton spin
asymmetry $A_{1,p}$ and the  structure function $g_1^p$ are presented. 
The data were collected during the year 2007.
We refer the reader to \cite{COMPASS_NIM} for the description of the 
160 GeV muon beam, the NH$_3$ polarized target and the COMPASS spectrometer.
DIS events are selected by cuts on the virtuality of the photon,
$Q^2 > 1$\,(GeV$/c)^2$, and its fractional energy, $0.1 < y < 0.9$.
The values of the spin-dependent structure function $g_1^p(x,Q^2)$  
have been combined with the published $g_1^d(x,Q^2)$ ones \cite{COMP_inc_d}
in order to evaluate the non-singlet  structure function $g_1^{NS}(x,Q^2)$.

We also present an evaluation of the helicity quark distributions
$\Delta u$, $\Delta d$, $\Delta\bar{u}$, $\Delta\bar{d}$ and $\Delta s$ 
(=$\Delta\bar{s}$) which were obtained in a combined analysis of 
inclusive and identified hadron asymmetries.
Both deuteron \cite{COMP_SIDIS_d} and proton data have been used.
In addition to the kinematic cuts mentioned above, for hadron tracks
coming from the primary vertex the cut $z>0.2$ is applied to select
the current fragmentation region. 
To avoid ambiguities between secondary muons and the scattered muon
we demand $z<0.85$.
Hadrons are identified as pions or kaons by the RICH detector
which limits their momenta to the range $10<p<50$\, GeV$/c$.

The total statistics of samples obtained with the proton target
for the inclusive, $\pi^+$ ($\pi^-$) and $K^+$ ($K^-$) events after 
all cuts is 92.5, 13.3 (11.8) and 3.9 (2.6) million events, respectively.


\section{Structure function $g_1^p$ and NLO QCD fit of $g_1^{NS}$}

\begin{figure*}[bt]
\centering
\epsfig{file=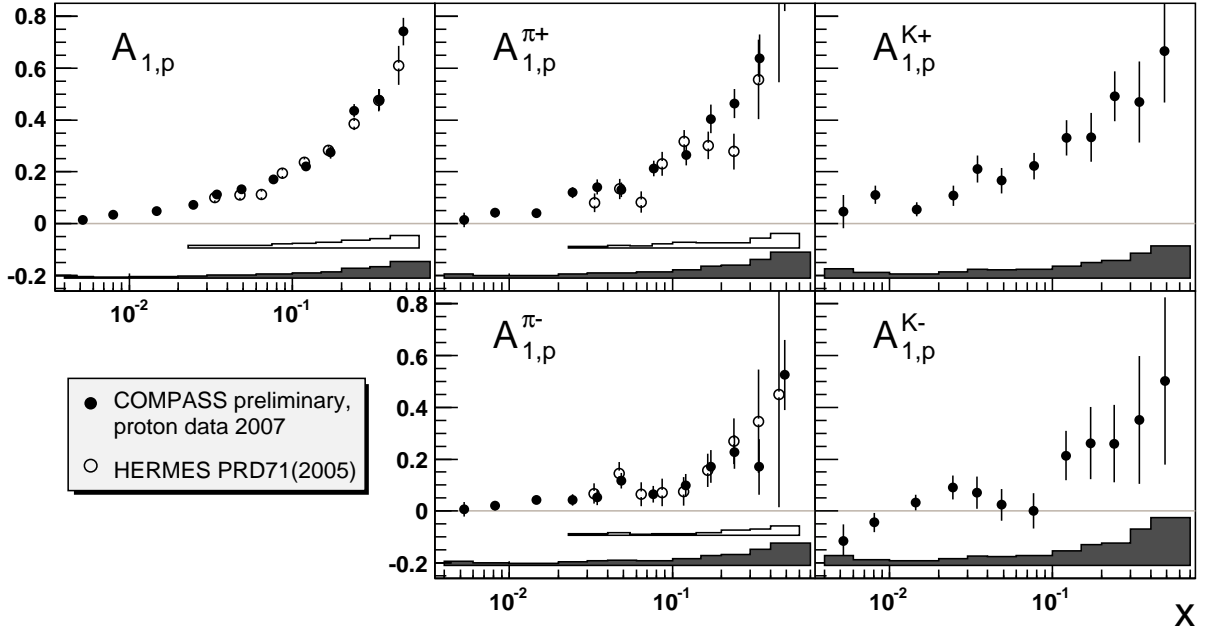,width=0.95\textwidth}
\caption{The double-spin asymmetries of COMPASS compared to
  results of HERMES \cite{HERMES}.
  Bands at bottom of each graph represent systematic uncertainties.
  Solid markers and bands correspond to COMPASS data.
  Open markers and bands are taken from HERMES.}
\label{fig:Asym}
\end{figure*}

The longitudinal virtual-photon proton asymmetry $A_{1,p}$ is defined
via the asymmetry of absorption cross sections of transverse photons as
\begin{equation}
A_{1,p} = \frac{\sigma_{1/2} - \sigma_{3/2}}{\sigma_{1/2} + \sigma_{3/2}},
\end{equation}
where the subscripts refer to the total  spin projection of the
$\gamma^{*}$-proton system.
The measured values of $A_{1,p}$ are shown as a function of $x$ in 
Fig.\,\ref{fig:Asym} in comparison with results from HERMES \cite{HERMES}.

\begin{figure}[tbh]
\epsfig{file=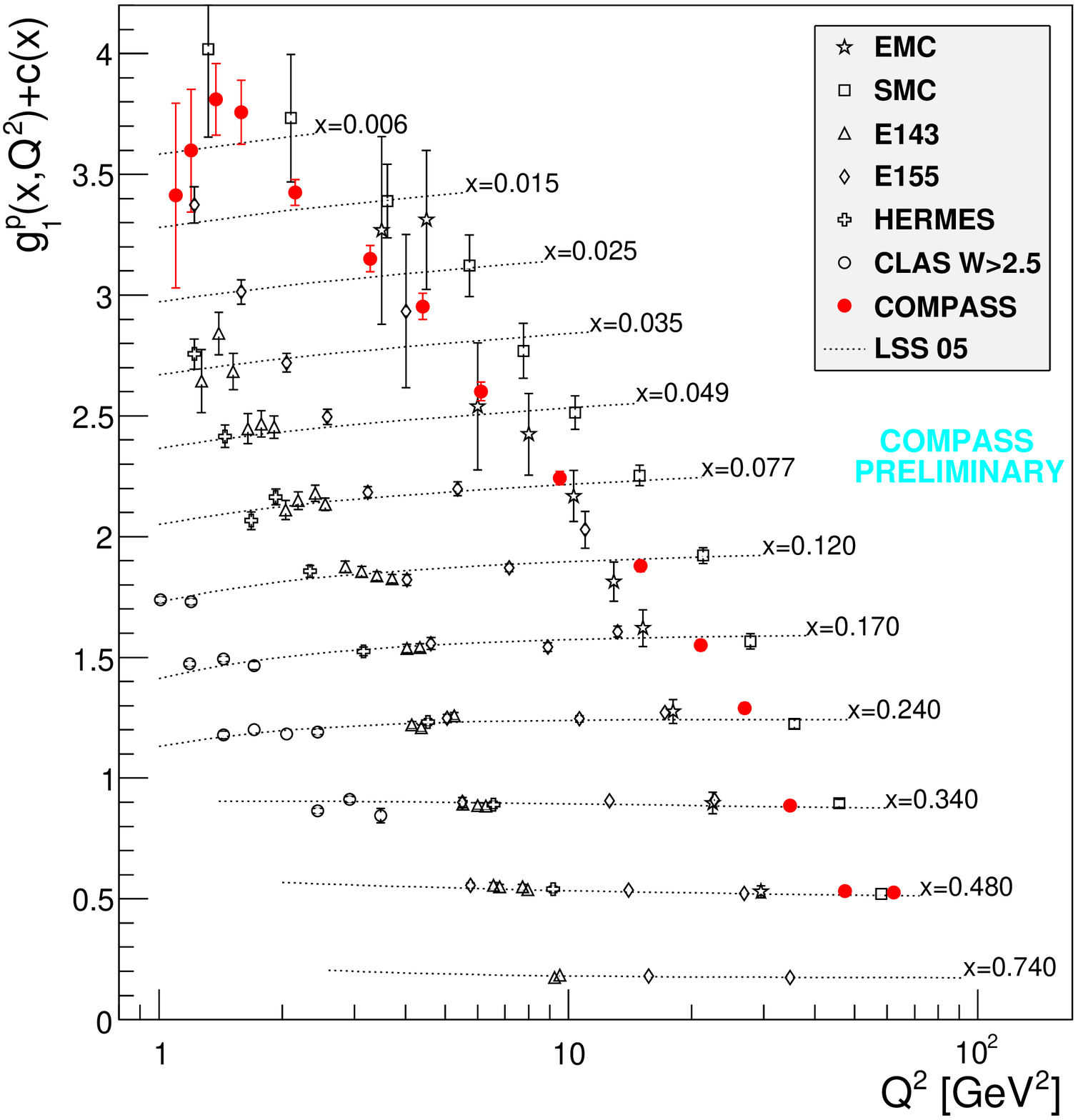,width=\linewidth}
\caption{$g_1^p$ as a function of $x$ and $Q^2$.
 To align points along curves corresponding to a fixed $x$,
 the LSS\,05 parametrization \cite{LSS05} has been used.
 Only statistical errors are shown.
 For the purpose of plotting, a constant $c(x)=0.28(11.6-i_x)$
 is added to each $g_1$ values, where $i_x$ is the number of 
 the $x$ bin ranging from $i_x=0$ ($x=0.006$) to $i_x=11$ ($x=0.74$).
}
\label{fig:g1p_all}
\end{figure}

The longitudinal spin structure function is obtained from $A_{1,p}$ as 
\begin{equation}
g_1^p = \frac{F_2^p}{2~ x~(1 + R)} A_1^p\,,
\end{equation}
where $F_2^p$ and $R$ are spin-independent structure functions.
$g_1^p$ as a function of $x$ and $Q^2$ is shown in Fig.\ref{fig:g1p_all}
where it is superposed to results of previous DIS experiments.

The availability of $g_1^d$ and $g_1^p$ data with good and 
comparable precision at low $x$ provides ideal conditions for 
a new evaluation of the non-singlet structure function  
\begin{equation}
\!\!\!\!\!\!
g_1^{NS}(x) = g_1^p(x) - g_1^n(x) = 2\Bigl[g_1^p(x)-\frac{g_1^d(x)}{1-3/2\omega_D} \Bigr],
~~~ 
\end{equation}
where $\omega_D$ is the probability of the D state
in the deuteron ($\omega_D=0.05\pm0.01$). 
An evaluation of the first moment of $g_1^{NS}(x)$
provides a new test of the Bjorken sum rule.   
This sum rule first derived using current algebra
is considered as a fundamental result of QCD.

The $Q^2$ dependence of  $g_1^{NS}$ is decoupled from the 
evolution of $\Delta \Sigma$ and $\Delta G$. 
Consequently a fit of the $Q^2$ evolution of $g_1^{NS}$ requires 
only a small number of parameters to describe the shape of $\Delta q_3(x)$ 
at some reference $Q^2$ and its integral, which is proportional to the
ratio $g_A/g_V$ of the axial and vector coupling constants.
The comparison of the fitted value of this ratio with the one
derived from neutron $\beta$ decay ($|g_A/g_V| = 1.2695\pm0.0029$ 
\cite{PDG})  provides a test  of the Bjorken sum rule, free of systematic
errors arising from uncertainties on $\Delta G$.

\begin{figure}[tb]
\epsfig{file=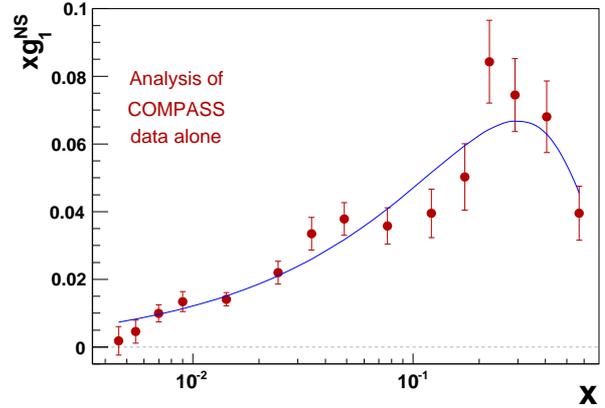,width=\linewidth}
\caption{Values of $g_1^{NS}(x)$ at $Q^2=3$ (GeV/$c)^2$. 
  The curve was obtained in the QCD NLO fit of the COMPASS
  data.}
\label{fig:g1NS}
\end{figure}

In the present analysis, $Q^2 = 3$ (GeV/c)$^2$ has been taken as
reference $Q^2$ and the following parametrization has been used 
for the isovector  distribution $\Delta q_3$:
\begin{equation}
\Delta q_3(x) = \frac{g_A}{g_V} 
\frac{x^{\alpha} (1 - x)^{\beta} }{\int_0^1 x^{\alpha} (1 - x)^{\beta} dx }
\end{equation}
The result of the fit is shown in Fig.\,\ref{fig:g1NS}.

The dominant systematic error is  due to the uncertainty of
5\% on the beam polarisation, which translates directly into a 5\% error
on $g_A/g_V$ (i.e.$\pm 0.065$). The other contributions due to $P_T$ and $f$
are $\pm 0.041$ for $A_{1,d}$ and $\pm 0.056$ for $A_1^p$. In total this
leads to a systematic error of $\pm 0.10$. The errors related to the fit
or to the evolution of the data to a common $Q^2$ are found to be negligible.
In particular, it was checked that the same value is obtained for $g_A/g_V$
when the reference $Q^2$ is 1.0, 3.0 or 10.0 (GeV/$c)^2$ although the shape
of $g_1^{NS}(x)$ becomes quite different.

Finally the following value have been obtained
\begin{equation}
| g_A/g_V | = 1.30\pm0.07(stat)\pm0.10(syst).
\label{gA_gV}                                                          
\end{equation}
It is in a perfect agreement with the one derived from neutron $\beta$ decay.


\section{LO QCD analysis for polarized quark densities}

As in our previous LO analysis \cite{COMP_SIDIS_d}, we assumed that
hadrons in the current fragmentation region are produced by independent
quark fragmentation, so that their spin asymmetries can be written 
in terms of parton distribution functions (PDFs) $q(x,Q^2)$, 
$\Delta q(x,Q^2)$ and fragmentation functions (FFs) $D_q^h(z,Q^2)$:
\begin{eqnarray} 
A^h(x,Q^2) = \frac{\sum_{q} e_q^2 \Delta q(x,Q^2) \int D_q^h(z,Q^2)dz } 
{\sum_{q} e_q^2 q(x,Q^2) \int D_q^h(z,Q^2) dz} \; . ~~
\label{eq:sidis}
\end{eqnarray}
In the present analysis we use the unpolarized PDFs from MRST \cite{MRST}
and the recent DSS parametrization of FFs at LO which was obtained 
from a combined analysis of inclusive pion and kaon production data 
from $e^+e^-$ annihilation, semi-inclusive DIS data from HERMES and 
proton-proton collider data \cite{DSS}.
The sensitivity of the obtained polarized PDFs on the choice of FFs
have been studied in \cite{COMP_SIDIS_d} where an alternative set
of FFs from the EMC experiment \cite{EMC_FF} has been used.

The 10 measured asymmetries (see Eq.\,\ref{eq:sidis}) form a linear 
system of equations with 5 unknowns $\Delta u$, $\Delta d$, $\Delta\bar{u}$,
$\Delta\bar{d}$ and $\Delta s$.
The system is solved by a  least-square fit independently in each $x$-bin.
Only statistical errors were used in the fit and correlations between 
asymmetries were taken into account.
A symmetry of the polarized strange sea is assumed, $\Delta s\equiv\Delta\bar{s}$.
In principle $\Delta{s}$ and $\Delta\bar{s}$ could both be extracted from 
the charged kaon asymmetries $A^{K+}_{p,d}$ and $A^{K-}_{p,d}$ 
but in view of the precision of the data, they are assumed to be equal.
All asymmetries are also assumed to be independent of $Q^2$.
In this way  the resulting PDFs are obtained at a common $Q^2$ fixed 
to 3\,(GeV/$c)^2$.

\begin{figure}[tbh]
\epsfig{file=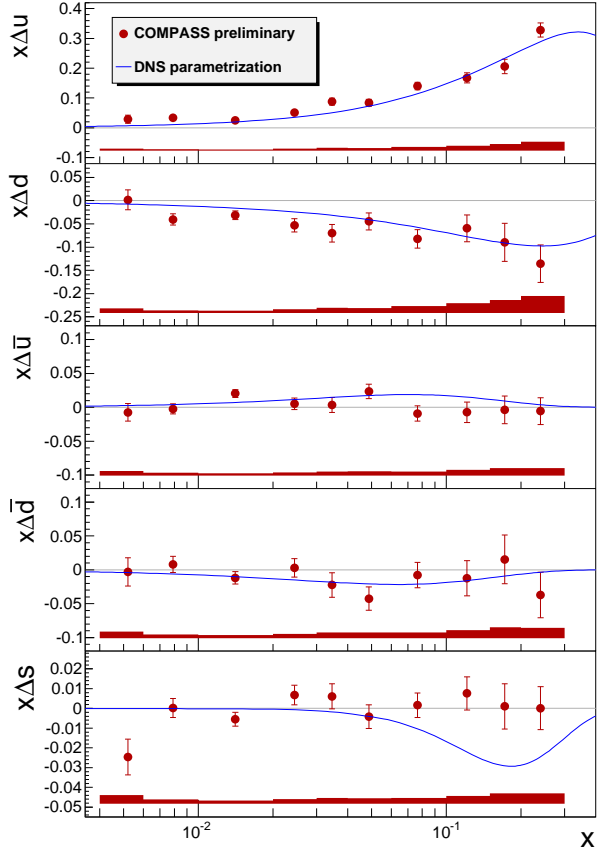,width=\linewidth}
\caption{The quark helicity distributions evaluated at common value
  $Q^2=3$\,(GeV/$c)^2$ as a function of $x$.
  Bands at bottom of each graph represent systematic uncertainties.
  For comparison the LO DNS parametrization \cite{DNS} is shown.}
\label{fig:PDFs}
\end{figure}

The quark helicity distributions, obtained with the DSS 
fragmentation functions, are shown as a function of $x$ 
with their  statistical and systematic errors  in Fig.\,\ref{fig:PDFs}.
The curves obtained with the LO DNS \cite{DNS} parametrization 
of polarized PDFs are also shown.

The $\Delta u$, $\Delta d$, $\Delta\bar{u}$, $\Delta\bar{d}$ 
curves from the DNS parametrization fit well our points.
However one can see a discrepancy in the $\Delta s$ graph.
The shape of the $x\Delta s$ curve of DNS is quite typical for 
QCD fits of $g_1(x,Q^2)$ data, with a  minimum in the medium $x$
region  $(x \approx 0.2)$.
The SIDIS measurements of COMPASS (it is also true for HERMES) do not
support this behavior, at least not with the DSS fragmentation functions.
More details on the distribution $\Delta s(x)$ are given in 
Ref.\,\cite{COMP_SIDIS_d}.
The present analysis including the 2007 proton data confirms 
the results obtained with the deuteron data and reduces significantly 
the statistical errors.

In the most recent global fit of polarized PDFs, DSSV \cite{DSSV}, a more
flexible parametrization for the strange quark distribution was adopted.
This fit includes HERMES data and therefore provides a positive
strange polarization at large $x$ and a negative one at small $x$.
A full comparison to our points obtained at LO is not possible
because the parametrization is provided at NLO only.


A sizable flavor asymmetry between the unpolarized up and down 
sea quark distributions (\,$\bar{u}(x)$$-$$\bar{d}(x)<0$\,) 
is a well established experimental fact since more than ten years
(see for instance \cite{MRST98} and references therein).  
It has  inspired a large theoretical activity and lead to 
various nonperturbative models which also  naturally predict 
a flavor asymmetry for the helicity densities of the light sea 
($\Delta \bar{u} -\Delta \bar{d} \ne 0$).
Most of the models, like Pauli-blocking models, instanton model, 
chiral quark soliton model, statistical models (see review \cite{Peng} 
and references therein),  predict a positive value of 
$\Delta\bar{u}$$-$$\Delta\bar{d}$.
The meson cloud models are the only ones which predict
a $\Delta\bar{u}$$-$$\Delta\bar{d}$ small in absolute value 
but with a negative sign.
Only experimental measurements can provide a test of these predictions.

\begin{figure}[tbh]
\epsfig{file=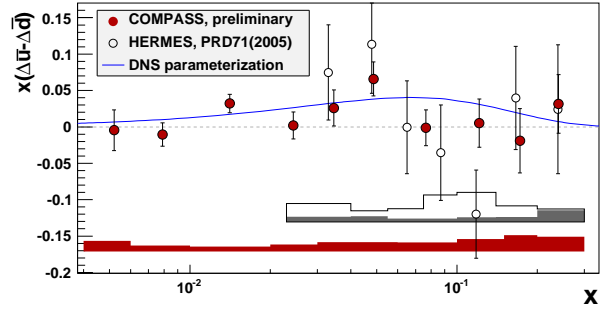,width=\linewidth}
\caption{The flavor asymmetry in the helicity densities of the light sea,
  $\Delta\bar{u}$$-$$\Delta\bar{d}$ evaluated at $Q^2=3$\,(GeV/$c)^2$.
  The data are compared with predictions of the DNS parametrization.
}
\label{fig:ub-db}
\end{figure}

The $x$ distribution of $\Delta\bar{u}$$-$$\Delta\bar{d}$ calculated
with the PDFs obtained in the LO fit described above is shown 
in Fig.\,\ref{fig:ub-db}.
For comparison the HERMES points are presented on the same plot.
A little excess of $\Delta\bar{u}$ with respect to $\Delta\bar{d}$ is seen.
The first moment of the distribution in the range of the measurements is
\begin{equation}
\!\!\!\!\!\!\!\!\!\!\!\!
\int_{0.004}^{0.3} \!\!(\Delta\bar{u}-\Delta\bar{d})dx =
0.052 \pm 0.035(stat) \pm 0.013(syst).~~~
\end{equation}
One should note that the systematics error does not include the uncertainty
from the FF parametrization since it was not provided by DSS authors.
A one-sigma effect in a narrower $x$ range is also claimed by HERMES:
$\int_{0.023}^{0.3} (\Delta\bar{u}-\Delta\bar{d})dx = 0.048 \pm 0.057(stat) \pm 0.028(syst)$.
The prediction of the DNS parametrization is shown in Fig.\,\ref{fig:ub-db}. 
One can notice a remarkable agreement of COMPASS measurements
with the DNS curve which is driven by the measurements of HERMES,
the only other experiment which measured asymmetries of identified
hadrons on a proton target.

\section{Summary}

A first COMPASS measurement of inclusive and identified hadron 
spin asymmetries with a proton target has been presented.
These asymmetries were included into a combined analysis together 
with the previously published deuteron ones. 
An evaluation of the non-singlet spin-dependent structure function              
$g_1^{NS}(x,Q^2)$ was presented.
The first moment of this function
confirms the validity of the Bjorken sum-rule with better than one $\sigma$
precision.

A LO evaluation of polarized quark densities was also presented.
The use of the proton in addition to deuteron data allowed 
to extract a full set of polarized PDFs $\Delta u$, $\Delta d$, 
$\Delta\bar{u}$, $\Delta\bar{d}$ and $\Delta s$ (=$\Delta\bar{s}$).
All sea quark densities are found to be compatible with
zero in the full range of measurements. 
However the discrepancy in a shape of $\Delta s(x)$ with the one
obtained in a typical global QCD fit was discovered.
A small excess of $\Delta\bar{u}$ with respect to $\Delta\bar{d}$ was observed.


\end{document}